\documentclass[10pt, conference, compsocconf, letterpaper]{IEEEtran}
\usepackage{graphicx}
\usepackage{algorithm}
\usepackage{algorithmic}

\newtheorem{defn}{Definition}

\def\IEEEbibitemsep{0pt plus .5pt}

\usepackage[caption=false,font=footnotesize]{subfig}

\begin{document}

\title{An empirical study of passive 802.11 Device Fingerprinting}

\author{
  \IEEEauthorblockN{Christoph Neumann, Olivier Heen, St\'ephane Onno}

  \IEEEauthorblockA{Technicolor Security and Content Protection Labs, Rennes, France
\\Email: \{christoph.neumann, olivier.heen, stephane.onno\}@technicolor.com}
}

\maketitle
\begin{abstract}
802.11 device fingerprinting is the action of characterizing a target device through its wireless traffic. This results in a signature that may be used for identification, network monitoring or intrusion detection.
The fingerprinting method can be active by sending traffic to the target device, or passive by just observing the traffic sent by the target device.
Many passive fingerprinting methods rely on the observation of one particular network feature, such as the rate switching behavior or the transmission pattern of probe requests.

In this work, we evaluate a set of global wireless network parameters with respect to their ability to identify 802.11 devices. We restrict ourselves to parameters that can be observed passively using a standard wireless card.
We evaluate these parameters for two different tests: i) the identification test that returns one single result being the closest match for the target device, and ii) the similarity test that returns a set of devices that are close to the target devices.
We find that the network parameters transmission time and frame inter-arrival time perform best in comparison to the other network parameters considered.
Finally, we focus on inter-arrival times, the most promising parameter for device identification, and show its dependency from several device characteristics such as the wireless card and driver but also running applications.
\end{abstract}


\section{Introduction}
Device fingerprinting is the action of gathering device information in order to characterize it.
This process generates a {\em signature}, also called a fingerprint, that describes the observed features of a device
in a compact form. If the generated signature is distinctive enough, it may be used to identify a device.

One application of 802.11 device fingerprinting is intrusion detection and more precisely
the prevention of Medium Access Control (MAC) address spoofing. 
MAC address spoofing is the action of taking the MAC address of another machine in order to 
benefit from its authorization. The prevention of MAC address spoofing is of importance in various scenarios.
Open wireless networks such as hot-spots often implement MAC address based access control in order
to guarantee that only legitimate client stations connect, e.g. the devices that purchased Internet access 
on an airport hot-spot.
An attacker can steal a legitimate device's session by spoofing the latter's MAC address.

Another application of fingerprinting is the detection of rogue access points (APs). 
Tools like AirSnarf 
or RawFakeAP 
enable an attacker to set-up a rogue AP, which result in client stations connecting to the fake AP instead of the genuine one.

Fingerprinting is also useful in key protected wireless networks (e.g. WPA2).
It can be used after a key-based authentication mechanism in order to control if only authorized devices are in the network.
Indeed wireless keys may leak as there are several normal situations where 
e.g. home users voluntarily give out their wireless key without renewing it afterwards, as
for instance when allowing a guest's laptop to access the home network.
While this scenario is both common and simple, it also endangers the home network;
the guest may abusively reconnect or the key may eventually leak from his laptop.
Tools like aircrack-ng exist that allow non-professional hackers to crack the (known to be insecure) WEP protocol.
Finally, commercial services, like wpacracker.com, exist which try to recover WPA keys. 

\paragraph{Contributions}
We investigate new passive fingerprinting candidates for 802.11 devices.
More precisely, we measure five network parameters that can be captured with a standard wireless card. 
Using a generic method to calculate a signature of a device, we compare the ability of 
each network parameter to characterize 802.11 devices.
We perform the similarity and the identification tests, which we evaluate using our own measurement data as well as public data obtained from large conference settings. We use detection window sizes of 5 minutes.
We find that the network parameters transmission time and frame inter-arrival time perform best in comparison to the other network parameters considered. Our evaluation renders a probability of correct classification that ranges from $79.4\%$ to $95.0\%$ for the transmission time and from $62.7\%$ to $93.1\%$ for frame inter-arrival times.
In the most difficult testing conditions, the wireless traffic of a conference, the inter-arrival time renders the best identification ratios. Up to $56.6\%$ of the devices could be uniquely identified with a false positive rate of 0.1.
Finally, we focus on inter-arrival times, the most promising parameter for device identification, and show its dependency from several device characteristics such as the device's wireless card and driver or running applications.

The remainder of this paper is organized as follows. We present related work 
in Section \ref{sec:related}. Section~\ref{sec:parameters} presents the different network parameters considered, and Section~\ref{sec:methodology} presents the proposed fingerprinting method.
Section~\ref{sec:eval} evaluates our method against several traces. 
We then present in Section~\ref{sec:factors} the  
802.11 features the generated signature depends on, and discuss possible 
attacks and applications in Section~\ref{sec:app}. Finally, we conclude
with Section~\ref{sec:conclusion}.

\section{Related Work}
\label{sec:related}

A first class of related work fingerprints wireless client stations by analyzing
implementation specificities of the network card and/or driver. Franklin et al. \cite{Franklin:Security06} 
characterize the drivers during the ``active scanning period''. 
This process is underspecified in the IEEE 802.11 standard regarding the frequency and order 
of sending probe requests. Therefore, each manufacturer implements its own algorithm and timers. 

Gopinath et al. \cite{Gopinath:WiNTECH} show that 802.11 cards exhibit very 
heterogeneous behavior due to implementation specificities. They test a set of 802.11 features such
as Random Back-off timers or Virtual Carrier Sensing and present their experimental results.
The observed heterogeneity in behavior may be used 
to fingerprint a card's vendor and model, but does not further analyze this aspect.

Bratus et al. \cite{Bratus:WISEC} propose an active method to fingerprint client stations as well as APs.
They send malformed or non-standard stimulus frames to the fingerprintee and apply decision trees on the observed response or behavior.
This yields a signature of the vendor/manufacturer.
Because it is active, an attacker can easily detect this technique.

Cache \cite{Cache:MasterThesis} proposes two methods for fingerprinting a device's network card and driver. 
The first one is active and uses the 802.11 association redirection mechanism. Even if well specified in 
the IEEE 802.11 standard, it is very loosely implemented in the tested wireless cards. As a consequence,  
each wireless card behaves differently during this phase which allows characterizing them. 
The second fingerprinting method of Cache is passive. It analyses the duration field values in 
802.11 data and management frames. Each wireless 
card computes the duration field in a slightly different way, which allows characterizing
the network card. 

Common to all above approaches is that they cannot differentiate between two devices using the same network card and driver. Therefore, those approaches may not be used for identifying individual devices.

Another class of related work allows fingerprinting individual APs.
Jana et al. \cite{Jana:Mobicom} calculate the clock skews of APs in order to identify them. 
The authors calculate clock skews by using the timestamps contained in Beacon frames 
emitted by the AP. 
Arackaparambil et al. \cite{Arackaparambil:Wisec} refine the above work and propose a new method yielding more 
precise clock measures. They also successfully spoof an AP, making it indistinguishable 
by the methods used by Jana et al. 

Loh et al. \cite{Loh:WiSec} fingerprint client stations, by observing 
probe requests. Stations send probe requests according to characteristic 
periodic patterns (see \cite{Franklin:Security06}).  The period itself is subject to slight variations. 
Far from being uniform, these variations can be clustered. 
With enough observation time, each cluster slowly derives, with a slope proportional to the time skew. 
This work is capable of uniquely identifying client stations; however, the requires more than one hour of traffic and
is only applicable to client stations.

In contrast to the above papers, Pang et al. \cite{Pang:Mobicom} discuss privacy implications in 802.11. 
Their paper highlights that users are not anonymous when using 802.11 as the protocol uses globally 
unique identifiers (the MAC address), which allow user tracking. 
Even if we suppose that this identifier is masked (e.g. by temporarily changing addresses) 
it is possible to track users by observing a set 
of parameters in the 802.11 protocol. 
The observed parameters are network destinations, network names 
advertised in 802.11 probes, 802.11 configuration options and broadcast frames' sizes.  
With encrypted traffic, three out of the four parameters still apply. 
The presented identification problem is close to our identification test.
To answer this test successfully, their fingerprinting technique requires traffic 
samples for each user that last at least one hour.


\section{802.11 network parameters}
\label{sec:parameters}
This section describes the network parameters we consider for fingerprinting.
We focus on network parameters that we can easily extract using a standard wireless card. We do not require the usage of expensive equipment such as software defined radios. Thus, the routine monitoring setup consists in a monitoring device that captures all 802.11 frames using a standard wireless card in monitoring mode on a specified 802.11 channel. 


We seek a fingeprinting method applicable on encrypted 802.11 traffic, making the fingerprinting method more universal and enabling fingerprinting devices from networks the monitoring device is not part of.

The fingerprinting method should not perturbate the network and should be hardly detectable by an attacker. As a result, our fingerprinting method is passive, i.e. the monitoring device does not generate any additional traffic.

We also require that the method relies on global network parameters, thus representing the traffic generated by a sender in general, rather than focusing on specific frames or features. In particular, it should be difficult for an adversary to deactivate or forge the considered network parameters. These considerations eliminate the option of extracting information from the 802.11 headers generated by the emitting station. The sender fills the fields in these headers, and the headers can thus be spoofed (e.g. using tools such as Scapy).

Finally, the fingerprinting method should be accurate, a property that we evaluate in the rest of this paper.

In light of the preceding requirements, we focus on information that we can extract solely from Radiotap \cite{radiotap.header} or Prism headers. 
The receiving wireless card driver generate these headers. An adversary that would like to change fields in these headers needs to change its behavior actually.

We consider the following network parameters, all being candidates for a fingerprinting method with the aforementioned requirements:
\begin{itemize}
\item{Transmission rate:} The 802.11 standard \cite{IEEE:80211} allows transmitting frames using a set of predifened rates
. Each sending wireless card chooses to transmit a given frame at a given rate.
Gopinath et al. \cite{Gopinath:WiNTECH} highlight that transmission rates distribution of wireless cards depends on the card's vendor. 
\item{Frame size:} The size of a 802.11 frame depends on the type of the frame, the fragmentation threshold, the version of IP\footnote{IPv4 addresses use 32 bits while IPv6 addresses use 128 bits. IP addresses are transported in 802.11 frames, thus changing the size of the frame.} or the applications generating the traffic. Pang et al. \cite{Pang:Mobicom} use broadcast frames size as an implicit identifier for wireless devices.
\item{Medium access time:} 
The time a wireless device waits after the medium as become idle and before sending its own frame.
Gopinath et al. \cite{Gopinath:WiNTECH} already noted that some device manufacturers implement the IEEE 802.11 standard very loosely with respect to the random backoff algorithm, which is one of the medium access mechanism of 802.11.
\item{Transmission time:} The transmission time is the time required to transmit a frame, thus the time between the start of reception and the end of reception of a frame.
\item{Frame inter-arrival time:} The frame inter-arrival time is the time interval between end of receptions of two consecutive frames. 
\end{itemize}

\section{Methodology}
\label{sec:methodology}
This section explains how we extract and evaluate signatures from the five network parameters considered in the previous section.

\subsection{Signature construction}

Signature calculation consists in generating several histograms, one histogram per frame type (e.g. Data frames, Probe Requests,\ldots). A histogram represents the frequencies of the values measured. Each histogram is weighted, which gives more or less importance to certain types of frame. 
We define the signature of device as the set of generated histograms generated by the device and their weights. 

We loose part of the information contained in a network trace by choosing histograms during signature calculation. Histograms may for instance eliminate characteristic patterns or periodic behaviors. Signal processing methods such as n-dimensional histograms, correlation functions or frequency analysis using Fourier transformations or Wavelet transformations may capture these behaviors. However, the subject of this work is not to find the most adequate signal processing method, but rather to highlight that some high level network parameters can achieve good detection ratios, even with a simple signature calculation method. We suspect that the network parameter that yields good performance with histograms will also yield the good results with more advanced signal processing methods. Section~\ref{sec:factors} provides an intuition on this latter assertion.
 
We now describe the signature generation process more formally. 
The sequence $f_0,\ldots,f_{n-1}$ of frames represents the network trace captured by the monitoring device.
$t_i$ denotes the time of end of reception of the frame $f_i$ (where $0 \le i \le n-1$), and frames are ordered in increasing reception time (i.e. $\forall i: t_{i-1} < t_{i}$). The sender $s_i$ sends the frame $f_i$.
For frames like ACK frames or clear-to-send frames \cite{IEEE:80211} the sender is unknown\footnote{ACK and CTS frames do not include a sender address or transmitting address field.}, thus $s_i = null$.

We calculate or extract the network parameter $p_i$ from the Radiotap or Prism header for each frame $f_i$.
Depending on the network parameter considered, $p_i$ may have different meanings.
Radiotap or Prism headers include the size $size_i$, the transmission rate $rate_i$ and the end of reception or the start of reception $t_{i}$ of a frame $f_i$. If the considered network parameter is the transmission rate, we have $p_i = rate_i$.  Similarly, $p_i=size_i$ if we consider frame sizes. We can also calculate the inter-arrival time $i_i =  t_{i} - t_{i-1}$, the transmission time $tt_i = size_i/rate_i$ and the medium access time $mtime_i = t_{i} - tt_{i-1}$. 

We add the measured or calculated parameter to the set $P^{ftype}(s_i)$. $P^{ftype}(s)$ denotes the set of values measured or calculated for frames of type $ftype$ for the sending device $s$. We denote $|P^{ftype}(s)|$ the number of observations for frames of type $ftype$ for device $s$.

\begin{figure}
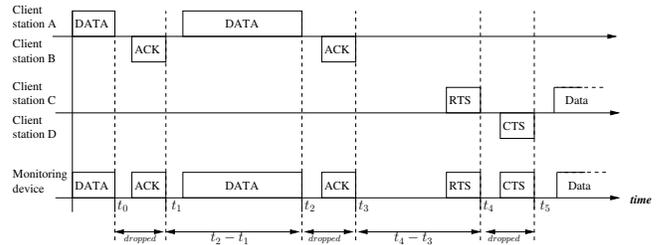

\begin{center}
\resizebox{\columnwidth}{!}{\input histogram.pdf_tex_t }
\caption{Measurement method example.}
\label{fig:histogram}
\end{center}
\end{figure}

Figure \ref{fig:histogram} illustrates our method. Client stations A, B, C and D use the same channel and send the frames as depicted. The monitoring device listens on the same channel and receives all frames of the emitting client stations. Thus, the sequence of frames $f_0,\ldots,f_5$ corresponds to the frame sequence DATA, ACK, DATA, ACK, RTS, CTS.
The first ACK frame $f_1$ has no explicit sender, $s_1 = null$. Thus, we drop the associated value $p_1$. 
Similarly, we drop the frames $f_3$ and $f_5$.
If we use transmission rate as a parameter,  we associate the value $rate_2$ to client station A, as frame $f_2$ is sent by station A.
We associate $rate_4$ to client station C. 
Thus, $P^{DATA}(A) = \{rate_2\}$ and $P^{RTS}(C) = \{rate_4\}$.
Similarly, if we use inter-arrival times as a parameter, we associate the interval $i_2 = t_2 - t_1$ to client station A.
We associate the interval $i_4 = t_4 - t_3$ to client station C. 
Thus, $P^{DATA}(A) = \{t_2 - t_1\}$ and $P^{RTS}(C) = \{t_4 - t_3\}$.

\begin{figure}
\begin{center}
\includegraphics[width=7cm]{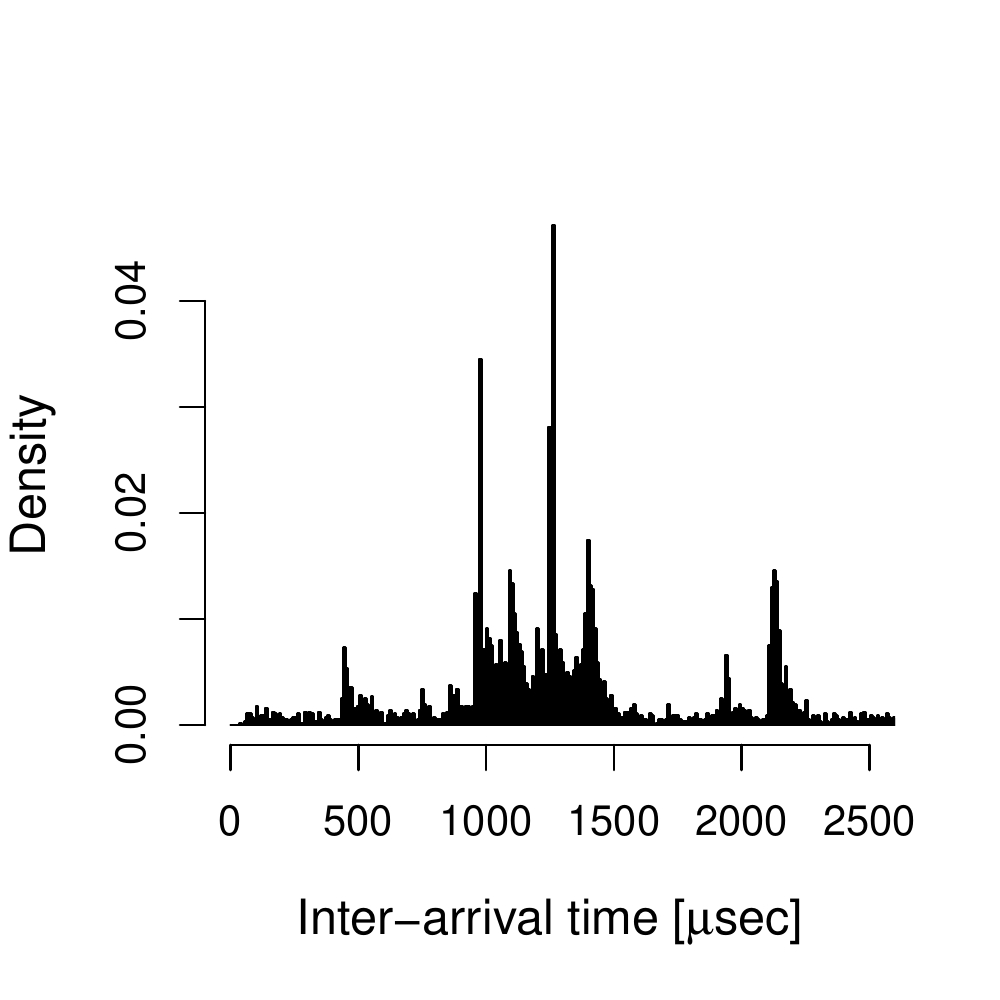}
\caption{Example inter-arrival time histogram}
\label{fig:examplehisto}
\end{center}
\end{figure}

Based on the above measurements we generate a histogram for each frame type and each emitting client station.
It is composed of bins $b_0,\ldots,b_{k-1}$. We denote $o^{ftype}_j$ (where $0 \le j \le k-1$) the number of observations in bin $b_j$.
We convert the histogram into a percentage frequency distribution, where the bin's $b_j$ percentage frequency   
is $P^{ftype}_j = o^{ftype}_j / |P^{ftype}(s)|$. The resulting histogram for a give frame type is $hist^{ftype}(s)=\{P^{ftype}_j| \forall j \in 0 \le j \le k-1\}$. Figure~\ref{fig:examplehisto} shows a resulting example histogram using inter-arrival times.

Finally, we define the signature $Sig$ of device $s$ as follows:
\begin{defn}[Device signature]
$$
Sig(s)= \{(weight^{ftype}(s),hist^{ftype}(s))| \forall ftype\}
$$
\end{defn}
The variable $weight^{ftype}$ weights the importance of a histogram for a given frame types.
For reference signature we choose that $$weight^{ftype}(s)=\frac{|P^{ftype}(s)|}{\sum_{ftype}|P^{ftype}(s)|}$$, thus the distribution of frame types is equivalent to the weight given to each frame type.


\subsection{Detection methodology and accuracy metrics}
\label{ssec:detection}
In general, fingerprinting methods have two phases: a learning phase which populates a reference database, and a detection phase which matches wireless devices against the reference database.

The reference database is built using a training dataset, which is in our case a wireless trace.
The reference database stores the signatures $Sig(r_i)$ of each wireless devices $r_i$, i.e. of each source address, appearing in the trace.
We suppose that no attacker polluted the training data.
The validity of this hypothesis is discussed in Section~\ref{sec:app}. 

In the detection phase, we analyze a second wireless trace, which we call validation dataset, 
and extract the signature $Sig(c_i)$ for all devices appearing in this trace. 
We call the devices of the validation dataset candidate devices. We compare
each candidate device's signature with all the signatures of the reference database using
the algorithm described in Section~\ref{sec:simil}. 
This yields a vector of similarities $<sim_1,sim_2,\ldots,sim_N>$, $sim_i$ corresponding 
to the similarity of the unknown candidate device's signature compared to the reference signature of device $r_i$.

For each candidate device, we are interested in resolving the following two tests: 
\begin{itemize}
\item \emph{Similarity:} The fingerprint algorithm returns a set of reference devices which signatures similarity
$sim_i$ is greater than a threshold $T$. Thus we are interested in knowing which reference devices are
similar to the candidate one.

\item \emph{Identification:} The second and more difficult test is interested in actually and uniquely identifying the candidate device.
To do so, we pick the reference device with the greatest similarity from the vector of similarities returned by the previous test. 

\end{itemize}

Regarding the similarity test we are interested in the {\em True Positive Rate} (TPR) and the {\em False Positive Rate} (FPR).
The TPR is the fraction of candidate wireless devices known to the reference database 
for which the returned set contains the actual device.
The FPR is the fraction of returned reference devices that do not match the actual candidate device.
In section~\ref{sec:eval}, we calculate the FPR and TPR as a function of the threshold $T$.

We plot a similarity curve which draws the TPR as a function of the FPR (see Section \ref{sec:eval}). We do not apply classical receiver operating characteristic (ROC) curves in this context, since we handle multiple classses (one class per reference device). In particular with the similarity curve, it is possible to have results in the lower right triangle of the plot.
Similarly to ROC curves, we also calculate the Area Under the Curve (AUC), which measures the global probability of correct classification.


We express the accuracy metric for identification test as an identification ratio. The identification ratio is the fraction of candidate wireless devices 
known to the reference database that the fingerprinting method correctly identifies. 
As with the similarity test, a false positive rate can be calculated for the identification test. 
For the identification test, the FPR is the fraction of candidate wireless devices that fingerprinting method mistakenly identifies as another 
device.

\subsection{Matching algorithm and similarity measures}
\label{sec:simil}

The algorithm below depicts how to match the signature $Sig(c)$ of a candidate $c$ against the reference database.
The algorithm returns a vector of similarities $<sim_1,sim_2,\ldots,sim_N>$, $sim_i$ corresponding 
to the similarity of the unknown candidate device's signature with the reference signature of device $r_i$.
We use the Cosine-similarity, as defined below, to calculate the similarity between two histograms.
We weight the resulting score with the frame type weight $weight^{ftype}(r_i)$ of the reference signature.

\begin{algorithm}[h!]
\begin{algorithmic}[]
{\footnotesize
\caption{{\bf Match}: Match the signature $Sig(c)$ of candidate $c$ against reference database}
\FORALL{$ftype \in Sig(c)$}
\STATE{Extract $hist^{ftype}(c)$ from $Sig(c)$}
\FORALL{references $r_i$ in reference database}
\STATE{Extract $hist^{ftype}(r_i)$ from $Sig(r_i)$}
\STATE{$sim^{ftype}_{i} = sim_{Cos}(hist^{ftype}(c),hist^{ftype}(r_i))$}
\STATE{$sim_{i} = sim_{i} + weight^{ftype}(r_i)*sim^{ftype}_{i}$}
\ENDFOR
\ENDFOR
\RETURN{$<sim_1,sim_2,\ldots,sim_N>$}
}
\end{algorithmic}
\label{matchalgo}
\end{algorithm}

Let $hist(r)=\{P_{r,j}|\forall j\}$ a reference histogram for device $r$. 
Let $hist(c)=\{P_{c,j}|\forall j\}$ a  candidate histogram for device $c$.

\begin{defn}[Cosine-similarity]
$$
sim_{Cos}(hist(c),hist(r)) = 
1 - \frac{\sum_{j=0}^{k-1} (P_{c,j}P_{r,j})}{ \sqrt{\sum_{j=0}^{k-1} P_{r,j}^2} \sqrt{\sum_{j=0}^{k-1} P_{c,j}^2}}
$$
\end{defn}

The Cosine-similarity is based on the Cosine-distance \cite{Cha:MATH08}.
The similarity equals $1$ if two signatures are exactly the same.
It yields $0$ when signatures have no intersection.
\section{Evaluation and Implementation}
\label{sec:eval}

\subsection{Wireless traces}

We evaluate our methodology against four different wireless traces. 
We use a publicly available 7 hours trace collected on August 19th 2008 at 11am
on one monitoring device during the 2008 Sigcomm conference \cite{umd-sigcomm2008-2009-03-02}. 
We consider two subsets of the Sigcomm trace: i) the entire 7 hours trace, that we call conference 1, and ii)
the 1st hour trace, extracted from the 7 hours trace, which we call conference 2.
We also generated two wireless traces ourselves. We generated the first one, called office 1, by capturing all wireless 
traffic on channel 6 during 7 hours in our office. We recorded the second one, called office 2, during one hour another day in the same setting. 
The conference traces are not encrypted (i.e. no WEP or WPA). 
The office traces are encrypted (WPA).

We evaluate our fingerprinting method by splitting each of the wireless traces described above
in two sets: i) a training dataset and ii) a validation dataset.
For the conference 1 and office 1 trace, the training dataset corresponds to the first hour of the 7 hour trace. The 6 remaining hours compose the validation dataset. For the two one hour traces, conference 2 and office 2,  we use
the first 20 minutes as a reference trace and the remaining 40 minutes as a training trace.
We use a detection window size of 5 minutes for the validation dataset. We match all candidate devices against the
reference database for each detection window. Using a minimum number of frames of 50 for generating
the signatures (see Section 4.3) we obtain training database sizes and number of candidates as shown in Table~\ref{tab:traces}.


\begin{table}
{\footnotesize
\begin{center}
\begin{tabular}{l|c|c|c|c}
 & Conf. 1& Conf. 2& Office 1 & Office 2 \\
 \hline
 Total duration & 7 hours & 1 hour& 7 hours & 1 hour \\
 Ref. duration & 1 hour & 20 min& 1 hour & 20 min \\
 Cand. duration & 6 hours & 40 min & 6 hours & 40 min \\
 Encryption & None & None& WPA & WPA \\
\# ref. devices & 188 & 97 & 158 & 120\\
\end{tabular}
\caption{Evaluation traces features.}
\label{tab:traces}
\end{center}
}
\end{table}

\subsection{Evaluation}

\subsubsection{Similarity}

\begin{figure}
\centerline{
\subfloat[Office 1]{
\includegraphics[width=7cm]{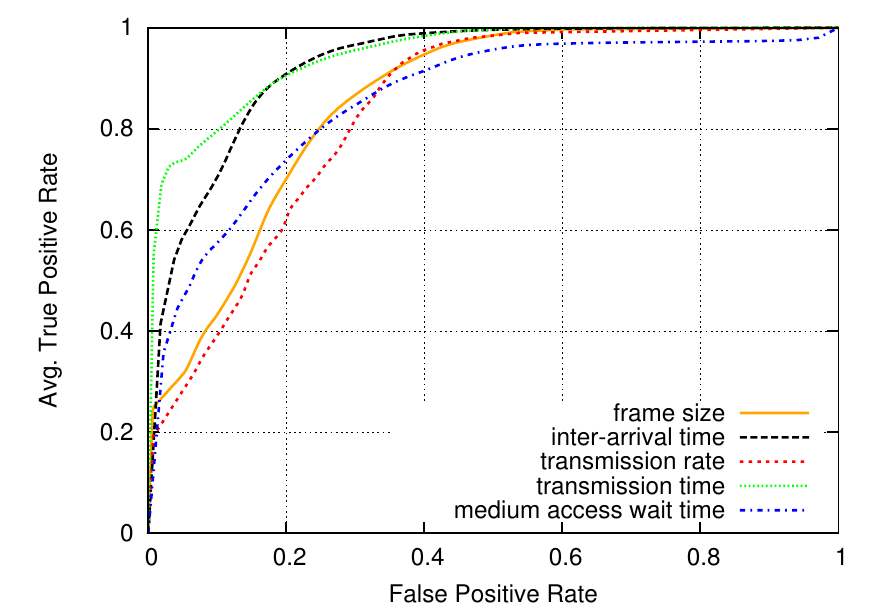}
}}
\hfil
\centerline{
\subfloat[Office 2]{
\includegraphics[width=7cm]{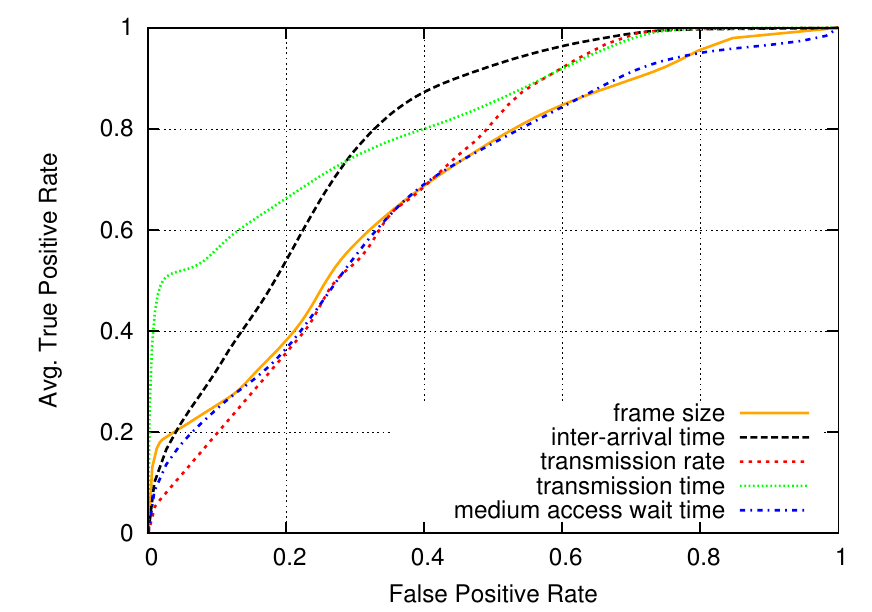}
}}
\hfil
\centerline{
\subfloat[Conference 1]{
\includegraphics[width=7cm]{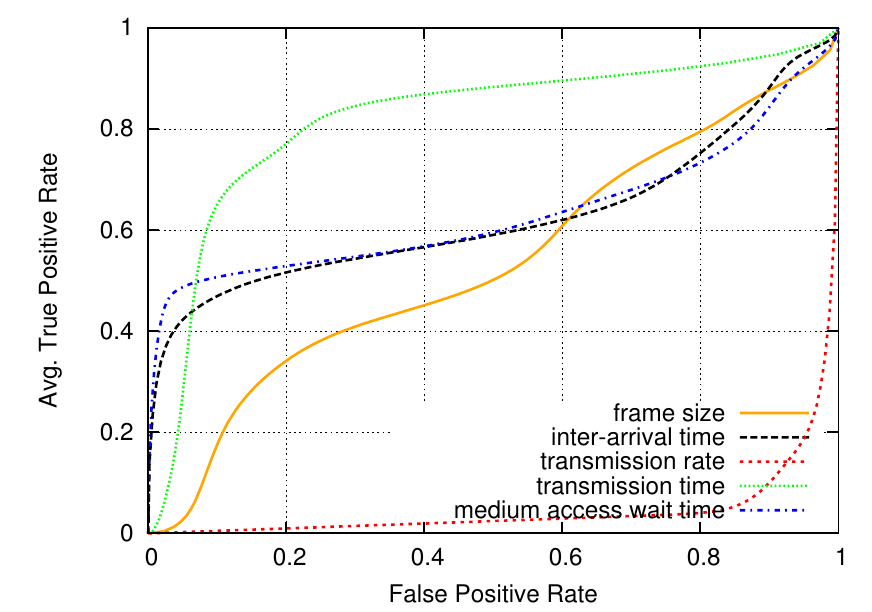}
\label{fig:sigcommprox1}
}}
\hfil
\centerline{
\subfloat[Conference 2]{
\includegraphics[width=7cm]{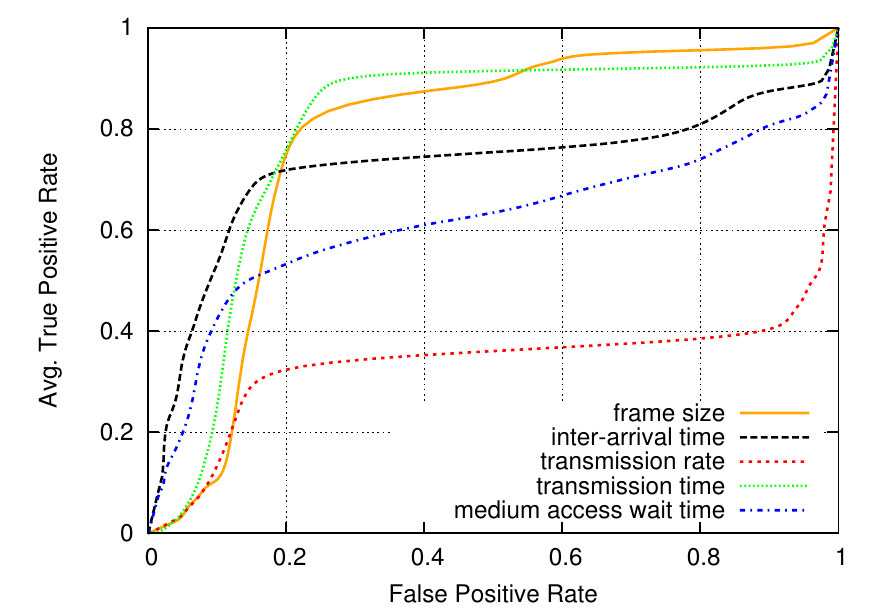}
\label{fig:sigcommprox2}
}}
\caption{Similarity curves (TPR vs. FPR) for office and conference traces. We do not apply  classical receiver operating characteristic (ROC) curves in this context, since we handle multiple classses (one class per reference device).}
\label{fig:SimFPRvsTPR}
\end{figure}

We first discuss the results for the similarity test as defined in Section~\ref{ssec:detection}.
We evaluate each of the network parameters using the classifier presented in Section~\ref{sec:simil} across several thresholds $T$. As $T$ decreases, the tolerated similarity between reference and candidate signature decreases, thus the FPR and TPR increases.

Figure~\ref{fig:SimFPRvsTPR} shows the similarity  curve for the traces office 1, office 2, conference 1 and conference 2. This curve draws the TPR as a function of the FPR. Table~\ref{tab:simauc} shows the Area Under the Curve (AUC) of these curves.

\begin{table}
{\footnotesize
\begin{center}
\begin{tabular}{l|c|c|c|c}
Network parameter  &  Conf. 1 & Conf. 2 &  Office 1 & Office 2 \\
\hline
Transmission rate& 4.0\%&33.5\%&83.7\%&70.6\%\\
Frame size&  53.4\%& 78.2\%&85.7\%&70.0\%\\
Medium access time&  63.4\%& 61.5\%&86.4\%&68.8\%\\
Transmission time&  80.7\%& 79.4\%&95.0\%&82.9\%\\
Inter-arrival time&  62.7\%& 72.5\%&93.7\%&80.1\%\\
\end{tabular}
\end{center}
}
\caption{Area Under the Curve (AUC) for the similarity test.}
\label{tab:simauc}
\end{table}

We observe that the transmission time generally outperforms all other network parameters, independently of the considered network trace.
If we consider the AUC, we can rank the network parameters in decreasing fingerprint accuracy as follows: transmission time, inter-arrival time, medium access time and transmission rate.
The transmission time achieves an AUC between $79.4\%$ and $95\%$. The inter-arrival time has similar results (with the exception of the conference 1 trace), with an AUC between $62.7\%$ and $93.7\%$. 
The medium access time achieves, the frame size and the transmission rate achieve an AUC between $61.5\%$ and $86.4\%$, $53.4\%$ and $86.7\%$ and $4.0\%$ and $83.7\%$ respectively.

A notable exception to above ranking is the behavior of the TPR for small FPRs in the conference setting. The two network parameters inter-arrival time and medium access time clearly outperform all other parameters.
With a FPR of 0.01, they yield a TPR between $7.8\%$ and $8.3\%$ for the short conference trace and a between $41\%$ and $45\%$ for the longer conference trace.
The transmission time only yields a TPR of $0.2\%$ for the short conference trace and of $13.6\%$ for the longer conference trace (FPR=0.01).

The conference setting is a more difficult setting for device fingerprinting than the office setting. The conference trace systematically yields lower AUC and TPRs, even with comparable reference database sizes and number of candidates. In addition, the relative difference of performance between the different network parameters becomes more important. Particularly the transmission rate has a poor fingerprint accuracy, due to the changing wireless conditions. In a conference setting, devices often change location which impacts the quality of the wireless signal and thus the detection ratio.

\subsubsection{Identification}

\begin{table}
{\footnotesize
\begin{center}
\begin{tabular}{l|c|c|c|c}
 Network parameter, FPR &  Conf. 1 & Conf. 2 &  Office 1 & Office 2 \\
\hline
Transmission rate, 0.01& 0\%&0.6\%&7.0\%&3.0\%\\
Transmission rate, 0.1& 0\%&7.5\%&12.9\%&7.0\%\\
Frame size, 0.01& 0\%& 0.2\%&18.4\%&13.8\%\\
Frame size, 0.1&  4.5\%& 2.5\%&33.9\%&20.4\%\\
Medium access time, 0.01&  22.7\%& 6.8\%&34.0\%&18.4\%\\
Medium access time, 0.1&  27.2\%& 28.1\%&41.0\%&21.1\%\\
Transmission time, 0.01&  0\%& 0\%&56.1\%&43.4\%\\
Transmission time, 0.1&  6.8\%& 5.8\%&60.5\%&50.5\%\\
Inter-arrival time, 0.01&  15.9\%& 6.4\%&48.0\%&21.5\%\\
Inter-arrival time, 0.1&  20.4\%& 32.2\%&56.7\%&27.5\%\\
\end{tabular}
\end{center}
}
\caption{Identification ratios.}
\label{tab:idratios}
\end{table}

We now present the results for the identification test shown in Table~\ref{tab:idratios}.

As with the similarity test, the transmission time outperforms the other network parameters in the office setting. 
With an FPR of 0.1, between 43.4\% and  60.5\% of the devices could be identified. With an FPR=0.01, between 50.5\% and 56.1\% of the devices could be identified. The remaining ranking in decreasing order of TPR is the inter-arrival time, the medium-access time and far behind with very poor results the frame size and the transmission rate.

In contrast to the office setting, the transmission time performs quite poorly in the conference traces. Instead, the inter-arrival time and the medium-access time outperform all other metrics. Using inter-arrival times, between 6.4\% and 15.9\% and between 20.4\% and 32.2\% of the devices could be identified with an FPR of 0.01 and 0.1 respectively. Using medium access times, between 6.8\% and 22.7\% and between 27.1\% and 28.1\% of the devices could be identified with an FPR of 0.01 and 0.1 respectively.

Such identification ratios might appear small. However, if we compare our results to the results obtained by Pang et al. \cite{Pang:Mobicom}, which analyzed a similar problem to our identification test, we achieve comparable results. For similar settings,
Pang et al. are able to detect 12\% to 52\% of users with a FPR of 0.1 and 5\% to 23\% of users with a FPR of 0.01.
In comparison, we could identify  27.1\% to 32.2\% and 6.8\% to 22.7\% of the devices with an FPR of 0.1 and 0.01 respectively.


In light of the different evaluation results, we consider only inter-arrival times in the rest of the paper.
This network parameter always appears in the top 3 network parameters (which are the transmission time, the inter-arrival time and the medium access time).
The inter-arrival time performs well in most setting. Even in the difficult setting of the conference trace the inter-arrival time yields good identification ratios. In contrast, the transmission time performs well in most scenarios but poorly in the most difficult setting of a conference.
Finally, the medium-access time has a similar behavior than the inter-arrival time but slightly underperforms in ''easy'' settings such as the office traces.

\subsection{Implementation}
\label{ssec:implem}

We have developed a tool in Python based on the {\tt pcap} library. It analyses standard pcap files as well as live traffic and extracts the different network parameters as described in Section~\ref{sec:methodology}. The tool also implements the fingerprinting methodology, i.e. the calculation of the device signatures, reference database, similarity measures and the calculation of accuracy metrics.




In our implementation we require that each training and candidate signature uses a
minimum number of 50 observations. Table~\ref{tab:traces} shows the resulting reference database sizes. 
$50$ observations correspond roughly to $50$ transmitted frames for the observed device.
The corresponding minimum observation time, i.e. the minimum time required to 
generate the signature, ranges from several seconds to several minutes.
It depends on the number of frames per second transmitted by the observed
device. For instance, for one of the devices of office 2 that did not generate much traffic this corresponds to 
30 seconds of traffic.
We also evaluated the performance with smaller thresholds, but we came to the conclusion 
that a minimum of 50 observations is a good compromise between the minimum time required to generate a signature and matching accuracy.

\section{Factors impacting the inter-arrival histogram shape}
\label{sec:factors}

Previous sections have shown that the frame inter-arrival time is the most promising network parameter 
with difficult monitoring conditions (typically in a conference setting) and for the more difficult 
test of unique device identification. This section discusses and demonstrates the different factors at various levels of the observed device that impact the frame inter-arrival time. 
Thus, this section gives an intuition why this network parameter performs better than the other network parameters proposed in Section~\ref{sec:parameters}.
This section also shows that inter-arrival times depend on other network parameters such as the medium access time and the transmission rate. Thus, this section also gives insights on the behavior of network devices regarding these parameters.

The inter-arrival time is composed of 
i) the transmission period and 
ii) the emitting client station's idle period.
Both periods have an impact on the signature value. We discuss different wireless device behaviors that impact either one or the other period.

\subsection{Wireless medium access methods}
\label{sec:accessmethods}
The 802.11 standard \cite{IEEE:80211} specifies mechanisms to avoid collisions among multiple devices competing for the same wireless medium. 
The wireless card or driver implement these mechanisms. Their effect is essentially expressed in the medium access wait time. 
Our evaluation (Section~\ref{sec:eval}) shows that the medium access wait time performs well for both the identification and the similarity test.

\subsubsection{Impact of random backoff}

\begin{figure}
\begin{center}
\includegraphics[width=3.9cm]{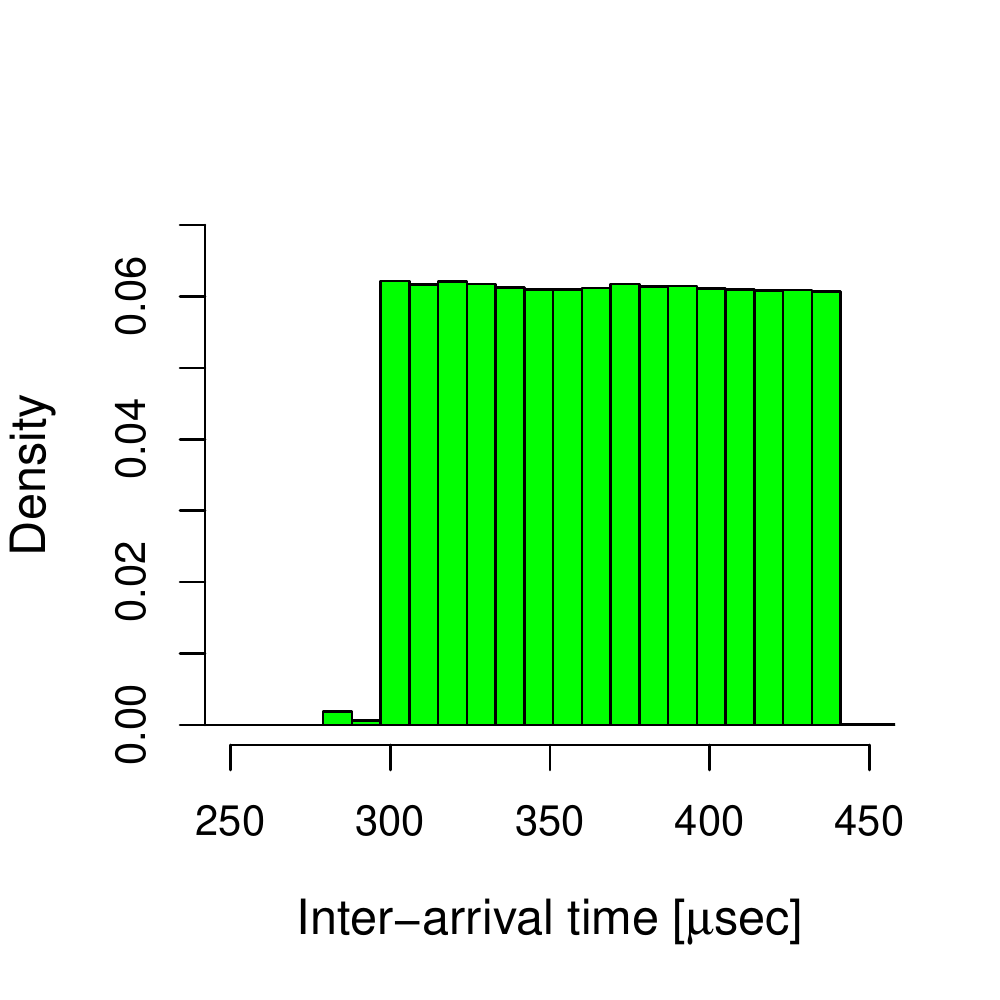}
\includegraphics[width=3.9cm]{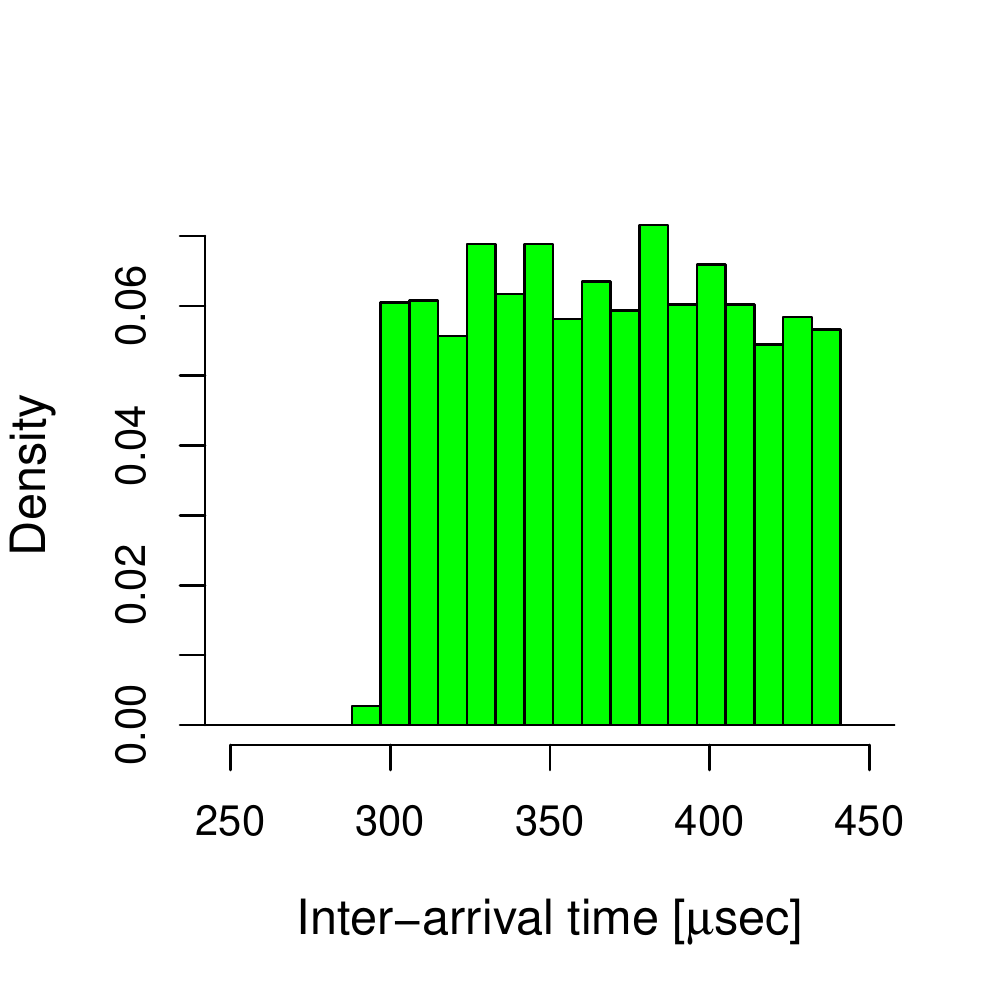}
\caption{Example inter-arrival histogram of two different wireless devices using different backoff implementations. Only data frames transmitted the first time (no retries) and sent at 54 Mbps are shown.}
\label{fig:backoff}
\end{center}
\end{figure}

The random backoff avoids frame transmission right after the medium is sensed idle. 
Instead, all client stations that would like to transmit frames should ensure that the medium is idle for a specified period, called DIFS, plus an additional random time, called backoff time before sending data.
Gopinath et al. \cite{Gopinath:WiNTECH} note that some device manufacturers implement the 802.11 standard very loosely. 
Berger et al. \cite{berger-lig2007} also note differences such as devices that systematically send frames during the first slot.

To evaluate the impact of these differences on the inter-arrival time histograms, we conduct the following experiment: 
We send a continuous UDP stream (using {\tt iperf}) from one wireless device placed within a Faraday cage. The Faraday cage minimized the impacts of external factors on the random backoff procedure.
In the second experiment, we just replace the sending device with a model from another manufacturer and sent the same UDP stream again.
We only analyze frames transmitted at 54 Mbps.
Figure~\ref{fig:backoff} shows the resulting histograms. 
We can notice that the first graph adds one small additional slot before the 16 slots defined by the standard. In addition, the distribution for the different slots is slightly different on both devices.
This indicates that the two devices implement the backoff mechanism differently.

\subsubsection{Impact of virtual carrier sensing}

\begin{figure}
\begin{center}
\subfloat[RTS mechanism deactivated.]{
\includegraphics[width=3.9cm]{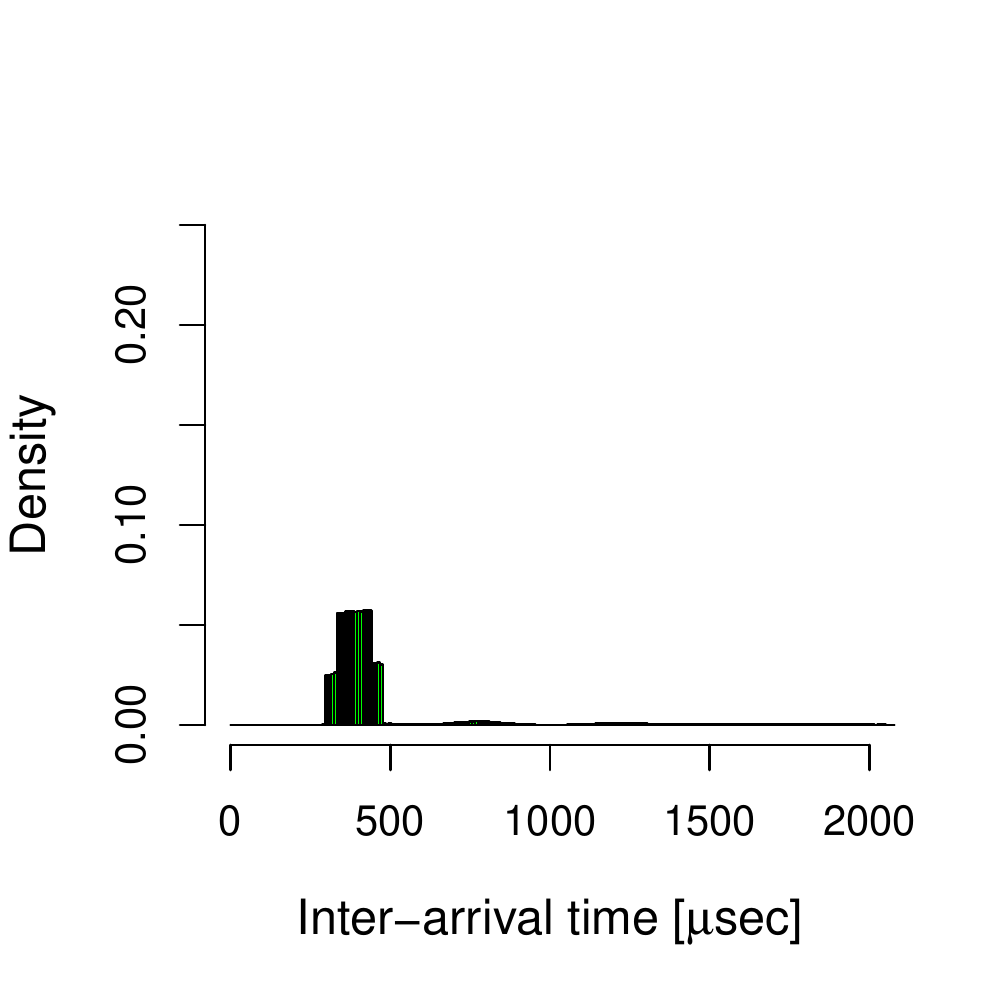}
}
\subfloat[RTS mechanism activated. RTS threshold set to 2000 bytes.]{
\includegraphics[width=3.9cm]{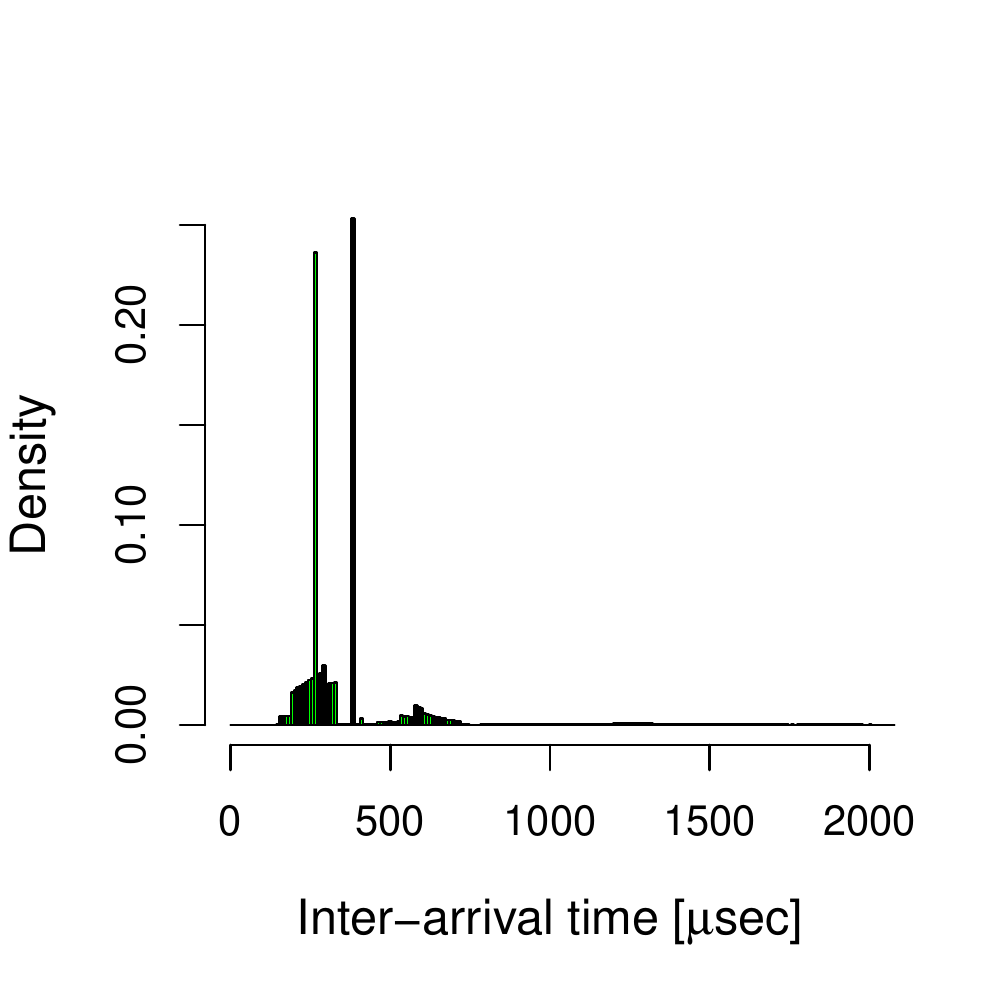}
}
\caption{Example inter-arrival histogram for the same device with different RTS settings.}
\label{fig:RTSthreshold_ALL}
\end{center}
\end{figure}

Virtual carrier sensing enables a client station to reserve the medium for a given amount of time (the contention-free-period).
The client station sends a {\em Request to Send} frame (RTS) and specifies the expected data transmission duration in this frame.
The destination device replies with a {\em Clear To Send} frame (CTS).
The client station can then transmit data frames during the reserved duration and all other stations are supposed to mute during the time specified. 
The idle period between two frames sent during contention-free-period is fixed to SIFS.
Wireless cards and drivers handle a so called RTS threshold, which is a value between $0$ and $2347$ bytes.

If the size of the data to be sent is greater than the RTS threshold, the virtual carrier sensing mechanism will be triggered.
Otherwise, the data frame will be sent using the random backoff mechanism.
In some wireless card driver implementations, the RTS threshold can be changed manually. 
In other ones, this threshold is hard-coded into the driver. 
Some devices do not implement this mechanism at all and exclusivley rely on the random backoff procedure.

To evaluate the impact of this mechanism on the inter-arrival time histograms, we conduct the following experiments: 
In a busy wireless network environment (our lab), a client station running under Linux sends a continuous UDP stream to a device connected by wire to an AP.
We use {\tt iperf} to generate the UDP stream.
We conduct the experiment twice using two distinctive RTS settings on the same sending client station:  
a) Virtual carrier sensing turned off and b) RTS threshold set to 2000 bytes.
Figure~\ref{fig:RTSthreshold_ALL} shows the resulting histograms. 
In Figure~\ref{fig:RTSthreshold_ALL}~a), all frames are sent after a random backoff mechanism.
In Figure~\ref{fig:RTSthreshold_ALL}~b), only RTS frames are sent after a random backoff mechanism, while data frames are sent during contention-free-period.

\subsection{Transmission rates}
\label{sec:transmissionrates}

\begin{figure}
\begin{center}
\subfloat[subfigcapmargin=tight][Device 1 signature]{
\includegraphics[width=3.9cm]{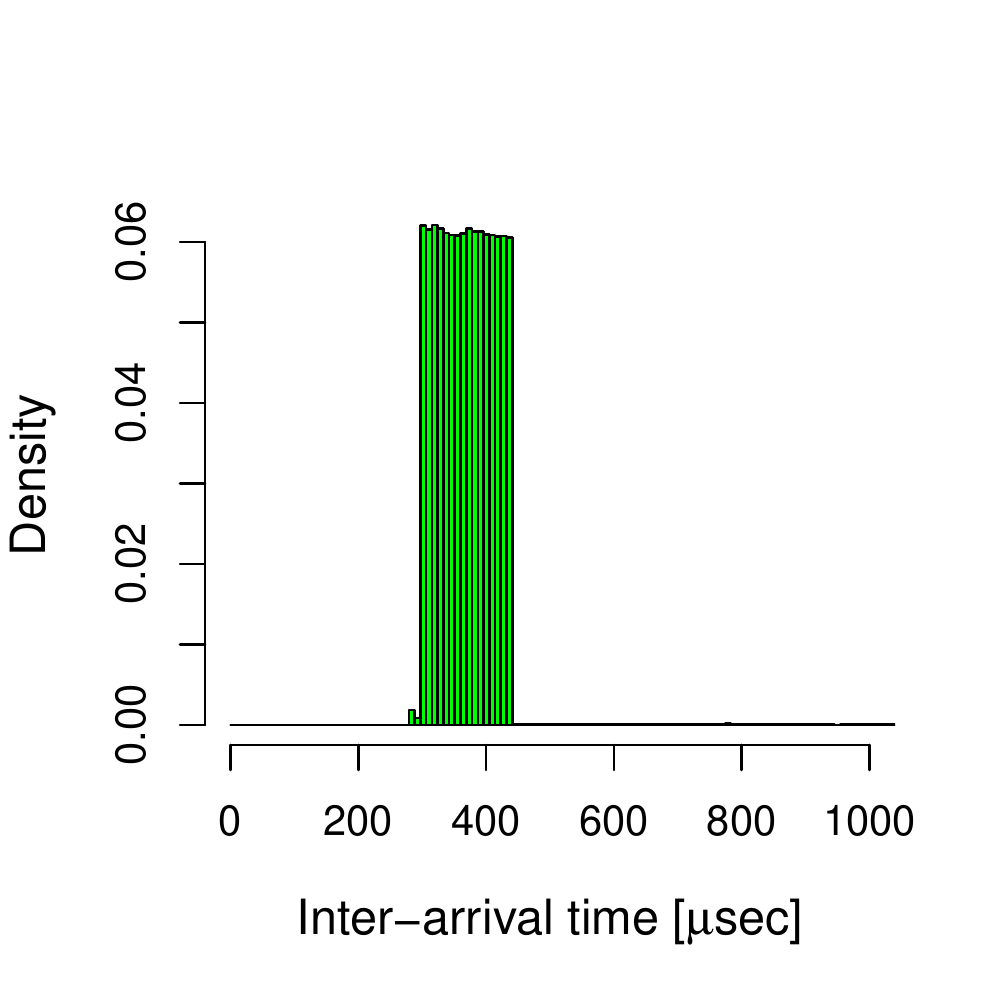}
}
\subfloat[subfigcapmargin=tight][Device 2 signature]{
\includegraphics[width=3.9cm]{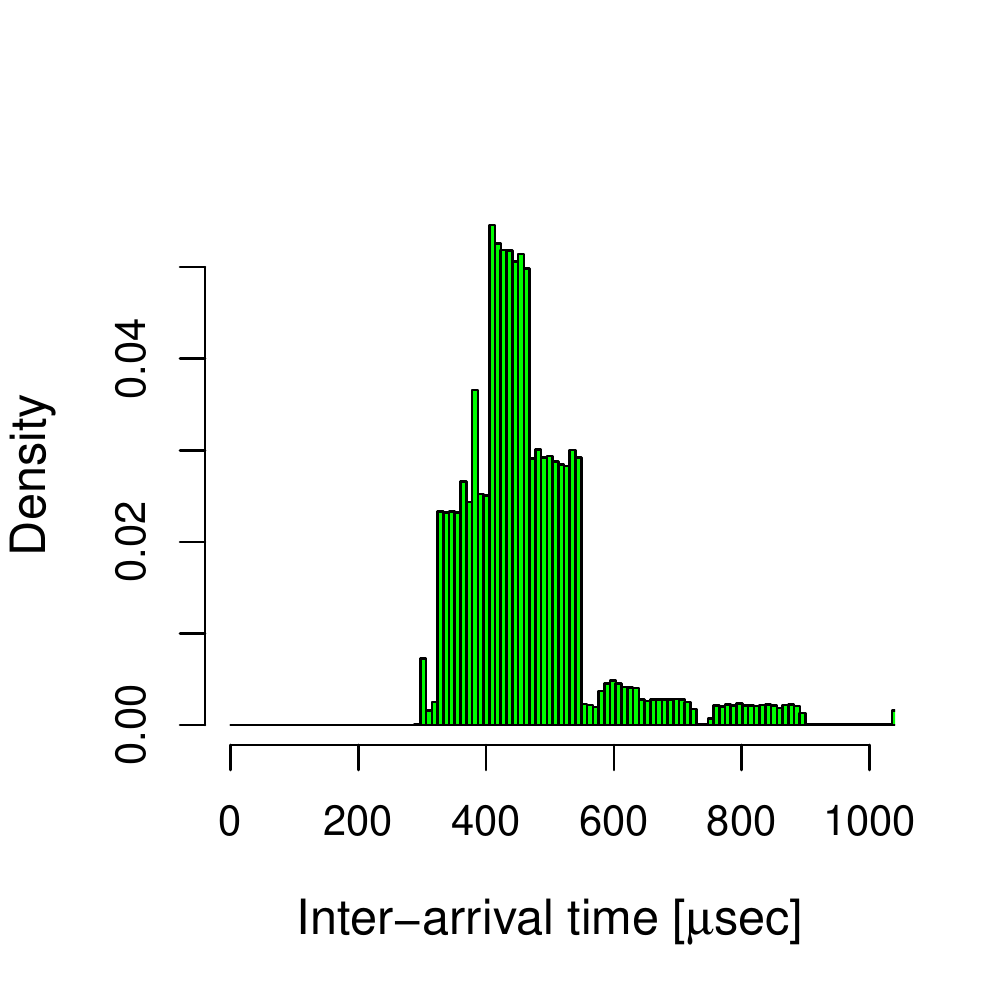}
}
\hfil
\subfloat[subfigcapmargin=tight][Device 1 transmission rate distribution]{
\includegraphics[width=3.9cm]{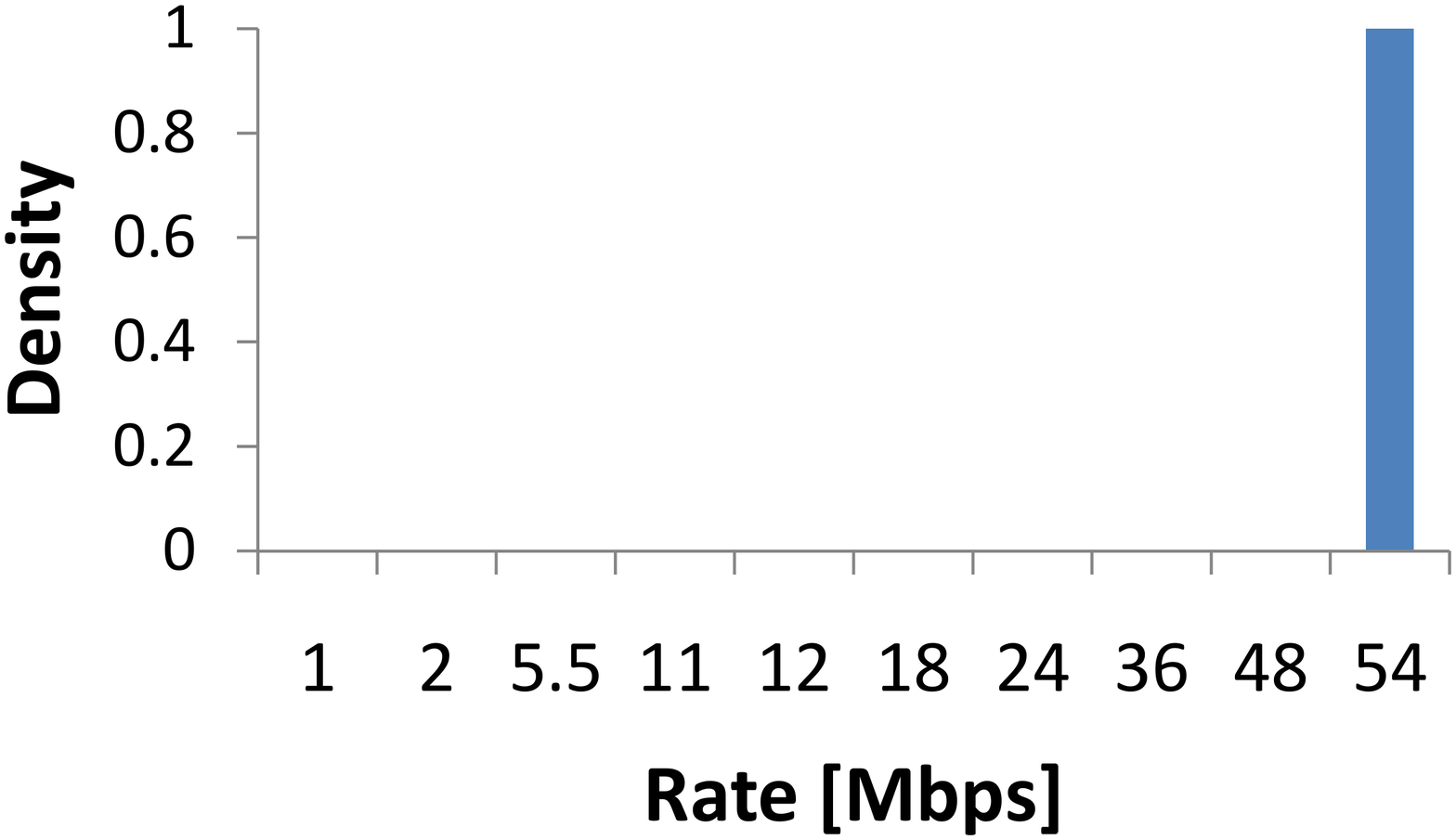}
}
\subfloat[subfigcapmargin=tight][Device 2 transmission rate distribution]{
\includegraphics[width=3.9cm]{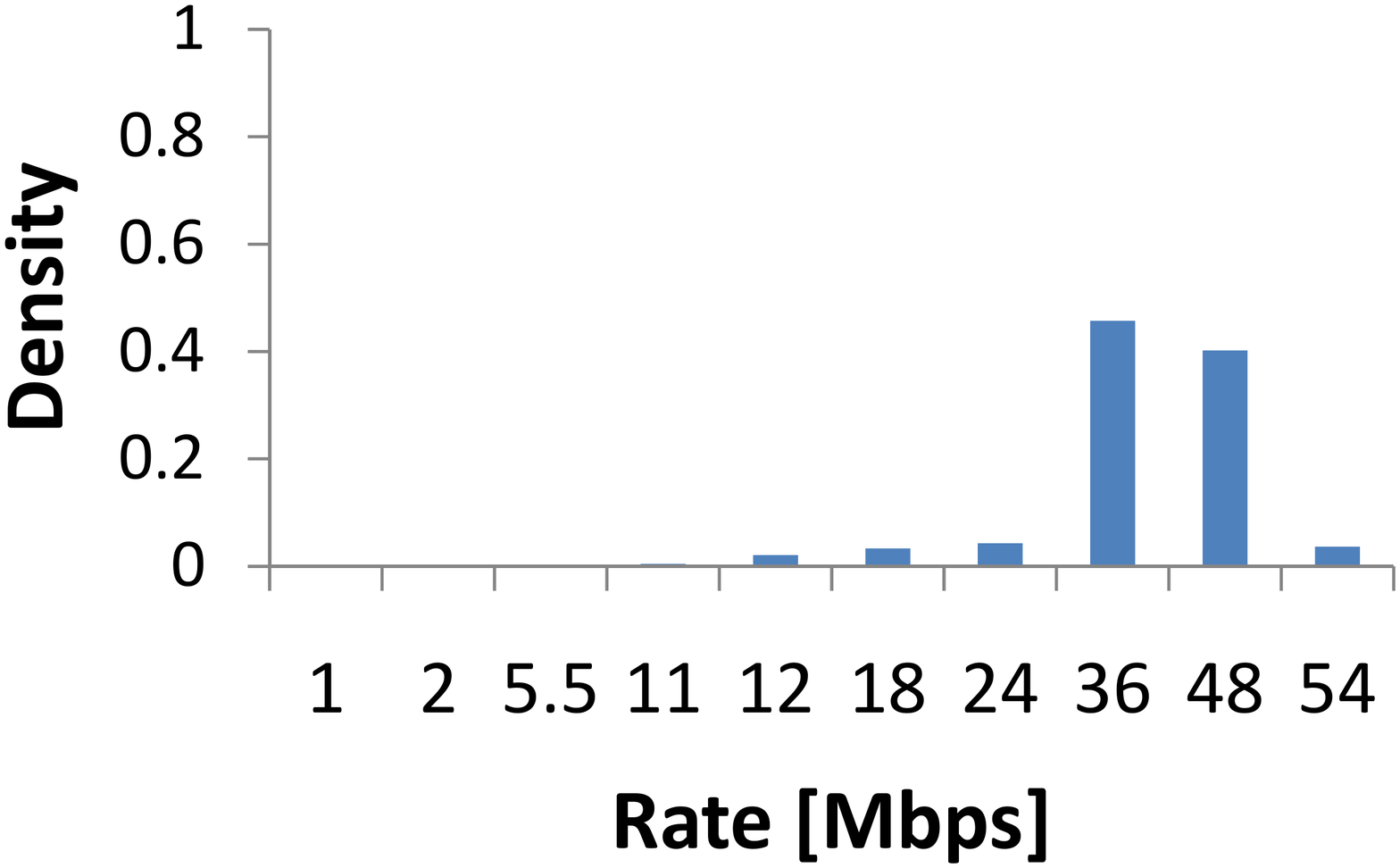}
}
\caption{Example signatures and transmission rate distributions of two different wireless devices using different transmission rates.}
\label{fig:transmissionrates}
\end{center}
\end{figure}

The transmission rates impact the time needed to transfer a frame. With inter-arrival times, we measure the frames end-of-reception time 
the effect of varying transmission rates is thus directly observable in our inter-arrival histograms. 

Our evaluation of the different network parameters (Section~\ref{sec:eval}) showed that transmission rates alone 
are not discriminant to measure device similarity and perform even more poorly to identify a device in a unique manner.
The inter-arrival time between two frames depends on the transmission rate of the frame. Thus, it is possible that the transmission rate has a negative impact on the performance of the inter-arrival time based fingerprints. 

Transmission rates can be quite discriminant in a controlled environment. Indeed, 
Gopinath et al. \cite{Gopinath:WiNTECH} highlight that data transfer rates distribution of wireless cards depends on the card's vendor. 
Similarly, \cite{conf/dsn/GaoCB10} shows that the rate switching behavior might be used to characterize a wireless access point.
The latter two papers made experiments in a very controlled environment.

We can illustrate the above behavior using the same Faraday cage experiment 
done for the random backoff timers. This time we include
all frames sent at various transmission rates in our measurements.
Figure~\ref{fig:transmissionrates} shows the resulting inter-arrival histograms and the distribution of used transmission rates.
We see that the second device changes its transmission rate more frequently. This yields a completely different histogram.




\subsection{Impact of network services}

\begin{figure}
\begin{center}
\subfloat[Netbook instance 1]{
\includegraphics[width=3.9cm]{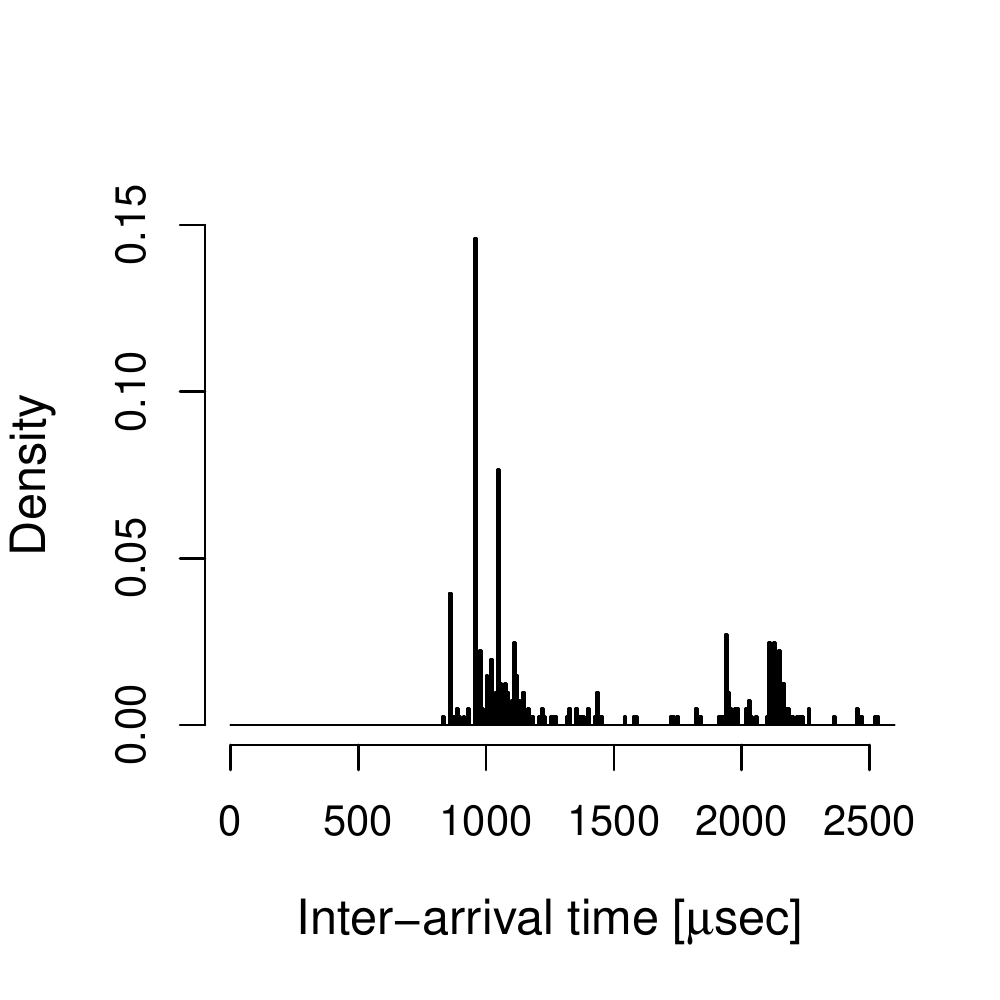}
\label{fig:applicative_peaks_red}
}
\subfloat[Netbook instance 2]{
\includegraphics[width=3.9cm]{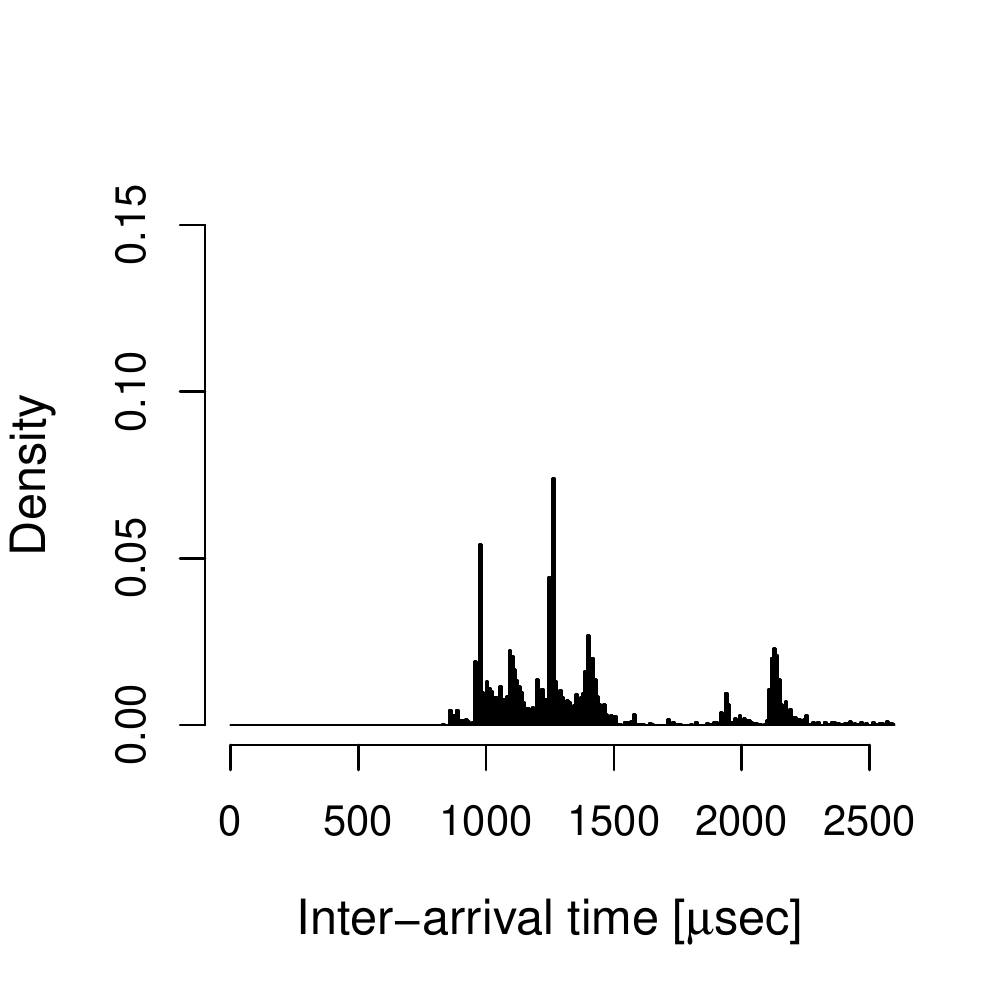}
\label{fig:applicative_peaks_black}
}
\caption{Example histogram based solely on data broadcast frames for two different devices with same model and same OS.}
\label{fig:applicative_peaks}
\end{center}
\end{figure}

Services and applications installed on a device influence the generated histograms. 
Applications generate the actual data traffic transferred over 802.11 and thus the device's traffic load. 
In addition, applications possibly generate very distinctive frames with specific frame sizes.
This is typically the case with network services such as Simple Service Discovery Protocol or Link-Local Multicast Name Resolution, running on the operating system. 
Our evaluation of the different network parameters (Section~\ref{sec:eval}) shows that frame sizes yield acceptable results for the
similarity test, however perform very poorly for the identification test.

Figure~\ref{fig:applicative_peaks} shows the impact of network services on inter-arrival histograms.
The figure only shows data broadcast frames. We generated the two histograms using two netbooks, same manufacturer, same model. Both netbooks ran the same operating system with the same updates. Both devices were active at the same time in the same wireless environment.  
Each device generates very distinctive peaks, eventhough both devices share the characteristics listed before. 
For instance in Figure~\ref{fig:applicative_peaks_black}, the protocols IGMPv3 and Link-Local Multicast Name Resolution generate the peaks at approx. 950 $\mu sec$ and 1200 $\mu sec$ respectively.
The device of Figure~\ref{fig:applicative_peaks_red} has another set of services running, yielding a different histogram.
Note that the latter figure also illustrates the results of Pang et al. \cite{Pang:Mobicom} which uses broadcast packets size as an implicit identifier.

\subsection{Other factors}

\begin{figure}
\begin{center}
\includegraphics[width=4cm]{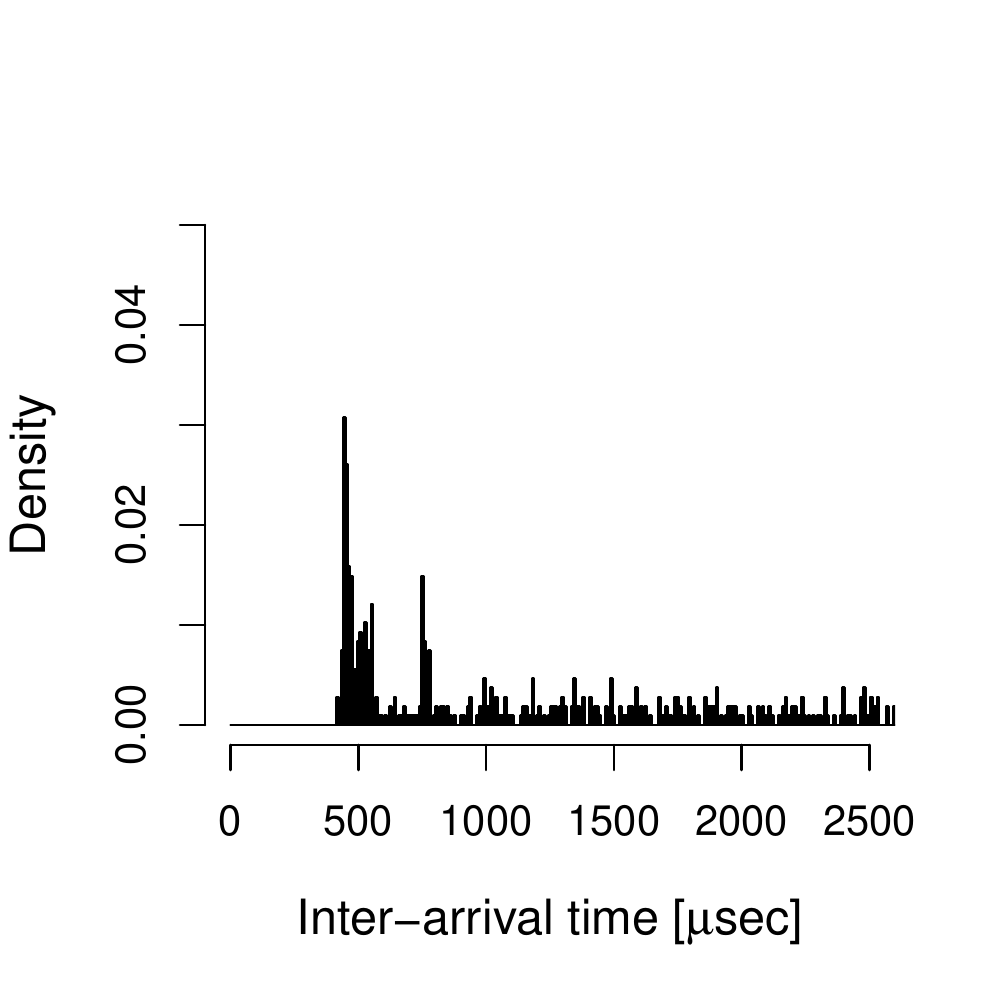}
\includegraphics[width=4cm]{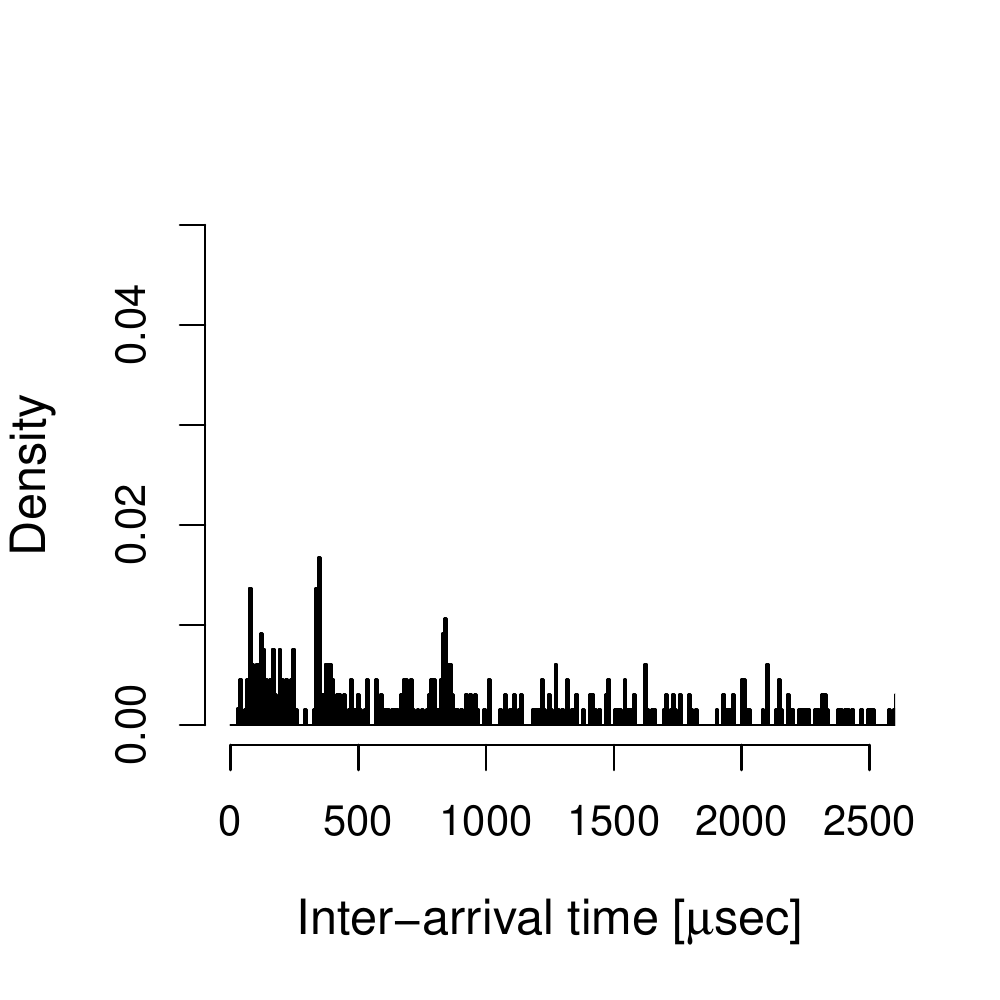}
\caption{Example histograms based solely on ``Data null function'' frames for two different wireless cards.}
\label{fig:histogrampowersavemode}
\end{center}
\end{figure}

Each wireless card supports a different 802.11 feature set or exhibits a specific behavior regarding some 802.11 features. 
For instance, Gopinath et al. \cite{Gopinath:WiNTECH} highlights that devices implement
the power save mode differently. 
We observe in our measurements that the power management feature generates additional traffic
in the histogram. We can isolate this traffic by observing the ``Data null function'' frames
which in most cases implement the power management feature. 
Figure~\ref{fig:histogrampowersavemode} shows two example histograms using two different wireless cards 
in the same wireless environment.
As we can see, the frequency distribution of this type of frames depends on the wireless card used.
Finally, several cards deactivate the power management feature under Linux. In this case, the traffic generated by this feature disappears.
Other similar features might be mentioned, such as Probe Requests frames
for which the literature already noted that driver specificities can be observed \cite{Franklin:Security06},
or similarly for Probe Responses in the case of AP's.



\section{Attacks and applications}
\label{sec:app}
In this section we discuss possible attacks against our fingerprinting method. We also discuss the applicability of our fingerprinting method in various contexts. We suppose that the signatures rely on inter-arrival times.

\subsection{Attacks}

\subsubsection{Forging a signature}
An attacker may try to fake the signature of a genuine device.
As many factors impact the signature (see Section \ref{sec:factors}) this can be a difficult task.
A casual attacker needs the same wireless equipment, driver and very similar driver and software configuration than the genuine device.
A more powerful attacker may record traffic of a genuine device and replay it, possibly live as in a relay attack.
The sensitivity to drivers and 802.11 parameters complicates the attacker task.
Indeed, the attacker must insert its own attacking traffic within the replayed traffic without modifying the signature.
This restricts the nature and quantity of the attacking traffic.

Another way of forging the signature is by learning and then trying to mimic the signature of a genuine device.
The attacker may send traffic at a constant transmission rate and vary the frame sizes for each frame type to reproduce the distribution of the histogram.
Some frame types, such as RTS or Probe requests, depend on the driver or chipset and thus require the attacker to change these features, which are more difficult to forge than application level data.

\subsubsection{Attacking the learning stage}
It is important that the learning stage is not polluted by the attacker.
While this is never guaranteed in real life situation, acceptable security is obtained if the learning stage starts 
on a moment chosen by the user or the administrator and not by the attacker.
Suppose that an AP would like to build a reference database. The learning stage might be initiated upon a user command (like pressing a button).
Then the AP learns the signatures of the allowed client stations during e.g. two minutes, which is sufficient with our fingerprinting method.

\subsubsection{Preventing fingerprinting}
802.11 communications are highly sensitive to denial of service attacks.
An attacker that is capable of performing a denial of service attack is also capable of preventing any fingerprinting activity.
A more subtle attacker may complicate the fingerprinting activity without blocking all the traffic.
Classically this consists in injecting fake frames using the MAC addresses of the genuine fingerprintees.
To the best of our knowledge, all wireless fingerprinting methods can be degraded by this attack.
Our fingerprinting method does not make an exception.

\subsection{Applications}

\subsubsection{Detecting fake client stations}
Access control based on the MAC address of a client station occurs in various contexts: enterprise network, hot-spots and home networks.
An attacker may want to spoof the MAC address of an authorized station in order to connect.
Our fingerprinting method is applicable in such context because forging an inter-arrival time signature is more difficult than just changing a MAC address.
An AP that routinely fingerprints its client stations against a reference database of allowed client stations would end-up noticing a non matching signature.
It can then warn the user or administrator that will react accordingly.


\subsubsection{Detecting rogue AP's}
\label{sec:appAP}



Fingerprinting can be used for detecting impersonation of an AP: 
A hot-spot operator may record and publish the signatures of valid APs. 
The client station accessing the hot-spot routinely fingerprints the AP and warns the user about mismatches.
In this case, the learning stage must be performed during a safe period: when receiving the AP from the vendor or during the installation of the hot-spot.
Our method is applicable to APs, if we ignore the data frames that the AP forwarded in lieu of another device.
Otherwise, applicative data generated by other devices would pollute the AP's signature.
This reduces the number of fingerprintable frames for an AP, but it is sufficient to
generate significant signatures.

\subsubsection{Localizing and tracking devices} 
Several authors propose fingerprinting as an approximate location mechanism \cite{AmbiantSensing,SurroundSense}.
In \cite{AmbiantSensing} a mobile device fingerprints its wireless environment and carries security decisions accordingly, 
like asking a password or not.
Our method is applicable in this case, because it produces signatures for both client stations and AP's, 
and because it requires only a few frames to generate signatures with a moving mobile device.

Finally, similar to \cite{Pang:Mobicom}, our work raises the question of privacy. Indeed the generated
signature may be used to trace a user's locations, even in cases where the device regularly 
changes its MAC address in order to stay anonymous.


\section{Conclusions}
\label{sec:conclusion}

We evaluated a set of global wireless network parameters with respect to their ability to identify 802.11 devices.
To do so, we defined a passive fingerprinting method that can be implemented with standard equipment. 
We considered that the network parameter frame inter-arrival time perform best in comparison to the other network parameters considered, in particular in the most difficult scenario of a conference setting.
Using this network parameter we are able to accurately identify 802.11 client stations and access points in a reasonable amount of time.

Whilst we gave some intuition about the ability to attack our approach
and explained to a certain extent which factors impact the 
shape of the histogram, future work should study these aspects in more detail. 
Especially the impact of applications and device updates on our fingerprinting method needs to be studied further.
Finally, future work should also investigate whether the fingerprinting method can be improved by combining several network parameters.

%
\bibliographystyle{abbrv}
\bibliography{biblio}  

\begin{thebibliography}{10}

\bibitem{radiotap.header}
{\em Radiotap}.
\newblock http://www.radiotap.org/.

\bibitem{IEEE:80211}
{\em {Wireless LAN Medium Access Control (MAC) and Physical Layer (PHY)
  Specifications}}.
\newblock IEEE 802.11 Standard, 1999.

\bibitem{Arackaparambil:Wisec}
C.~Arackaparambil, S.~Bratus, A.~Shubina, and D.~Kotz.
\newblock {On the Reliability of Wireless Fingerprinting using Clock Skews}.
\newblock In {\em ACM WiSec}, 2010.

\bibitem{SurroundSense}
M.~Azizyan, I.~Constandache, and R.~R. Choudhury.
\newblock {SurroundSense: Mobile Phone Localization via Ambience
  Fingerprinting}.
\newblock In {\em ACM MobiCom}, 2009.

\bibitem{berger-lig2007}
G.~Berger-Sabbatel, Y.~Grunenberger, M.~Heusse, F.~Rousseau, and A.~Duda.
\newblock {Interarrival Histograms : A Method for Measuring Transmission Delays
  in 802.11 WLANs}.
\newblock Research report, LIG lab, Grenoble, France, 2007.

\bibitem{Bratus:WISEC}
S.~Bratus, C.~Cornelius, D.~Kotz, and D.~Peebles.
\newblock Active behavioral fingerprinting of wireless devices.
\newblock In {\em ACM WiSec}, 2008.

\bibitem{Cache:MasterThesis}
J.~Cache.
\newblock {\em {Fingerprinting 802.11 Devices}}.
\newblock Master Thesis, 2006.

\bibitem{Cha:MATH08}
S.-H. Cha.
\newblock {Taxonomy of Nominal Type Histogram Distance Measures}.
\newblock In {\em MATH}, 2008.

\bibitem{Franklin:Security06}
J.~Franklin, D.~McCoy, P.~Tabriz, V.~Neagoe, J.~V. Randwyk, and D.~Sicker.
\newblock {Passive Data Link Layer 802.11 Wireless Device Driver
  Fingerprinting}.
\newblock In {\em Usenix Security}, 2006.

\bibitem{conf/dsn/GaoCB10}
K.~Gao, C.~L. Corbett, and R.~A. Beyah.
\newblock A passive approach to wireless device fingerprinting.
\newblock In {\em DSN}. IEEE, 2010.

\bibitem{Gopinath:WiNTECH}
K.~Gopinath, P.~Bhagwat, and K.~Gopinath.
\newblock {An Empirical Analysis of Heterogeneity in IEEE 802.11 MAC Protocol
  Implementations and its Implications}.
\newblock In {\em ACM WiNTECH}, 2006.

\bibitem{Jana:Mobicom}
S.~Jana and S.~K. Kasera.
\newblock {On Fast and Accurate Detection of Unauthorized Wireless Access
  Points Using Clock Skews}.
\newblock In {\em ACM MobiCom}, 2008.

\bibitem{AmbiantSensing}
N.~Kasuya, T.~Miyaki, and J.~Rekimoto.
\newblock {Activity-based Authentication by Ambient Wi-Fi Fingerprint Sensing}.
\newblock In {\em ACM MobiCom}, 2009.

\bibitem{Loh:WiSec}
D.~C.~C. Loh, C.~Y. Cho, C.~P. Tan, and R.~S. Lee.
\newblock {Identifying Unique Devices through Wireless Fingerprinting}.
\newblock In {\em ACM WiSec}, 2008.

\bibitem{Pang:Mobicom}
J.~Pang, B.~Greenstein, R.~Gummadi, S.~Seshan, and D.~Wetherall.
\newblock {802.11 User Fingerprinting}.
\newblock In {\em ACM MobiCom}, 2007.

\bibitem{umd-sigcomm2008-2009-03-02}
A.~Schulman, D.~Levin, and N.~Spring.
\newblock {CRAWDAD} data set umd/sigcomm2008 (v. 2009-03-02).
\newblock Downloaded from http://crawdad.cs.dartmouth.edu/umd/sigcomm2008,
  2009.

\end{thebibliography}
\end{document}